\documentclass[prb,reprint,superscriptaddress,
amsmath,amssymb,aps,]{revtex4-2}
\usepackage{array}
\usepackage{amsmath}
\usepackage{graphicx}
\usepackage{dcolumn}
\usepackage{bm}
\usepackage{float}
\usepackage{hyperref}
\usepackage{physics}
\hypersetup{colorlinks,linkcolor={blue},citecolor={blue},urlcolor={blue}}  
\usepackage{xtab,afterpage,longtable}
\usepackage[utf8]{inputenc}
\DeclareUnicodeCharacter{03C3}{\ensuremath{\sigma}}
\makeatletter
\def\LT@LR@e{\LTleft\z@   \LTright\z@}%
\makeatother

\usepackage[utf8]{inputenc}
\usepackage{textcomp}
\DeclareUnicodeCharacter{2212}{\textminus}

\begin{document}

\preprint{APS/123-QED}

\title{Interplay of non-Hermitian skin effect and electronic correlations in the non-Hermitian Hubbard model via Real-space dynamical mean field theory}
\author{Chakradhar Rangi}
\email{crangi1@lsu.edu}
\affiliation{Department of Physics and Astronomy, Louisiana State University, Baton Rouge, LA 70803, USA}
\author{Juana Moreno}
\affiliation{Department of Physics and Astronomy, Louisiana State University, Baton Rouge, LA 70803, USA}
\affiliation{Center for Computation and Technology, Louisiana State University, Baton Rouge, LA 70803, USA}
\author{Ka-Ming Tam}
\affiliation{Department of Physics and Astronomy, Louisiana State University, Baton Rouge, LA 70803, USA}
\affiliation{Center for Computation and Technology, Louisiana State University, Baton Rouge, LA 70803, USA}
\date{\today}
\begin{abstract}
Non-Hermitian quantum systems, characterized by their ability to model open systems with gain and loss, have unveiled striking phenomena such as the non-Hermitian skin effect (NHSE), where eigenstates localize at boundaries under open boundary conditions. While extensively studied in non-interacting systems, the interplay between NHSE and strong electron correlations remains largely unexplored. Here, we investigate the non-Hermitian Hubbard model with asymmetric hopping, employing real-space dynamical mean-field theory (R-DMFT), a novel extension to such non-Hermitian correlated models—to capture both local correlations and spatial inhomogeneities. By analyzing the end-to-end Green's function as probes of directional amplification, we demonstrate that strong correlations can suppress the skin effect, leading to a crossover from boundary-dominated to correlation-driven dynamics. Our systematic study reveals that correlations dominate at small to intermediate strength of asymmetric hopping, inducing exponential decay in the end-to-end Green's function, but higher strength of asymmetric hopping can restore amplification even at strong interaction. These results illuminate the tunable interplay between correlations and non-Hermitian physics, suggesting avenues for engineering non-reciprocal transport in correlated open quantum systems.
\end{abstract}\maketitle

\section{Introduction}

The advent of non-Hermitian quantum mechanics has unveiled a rich tapestry of quantum phenomena in open systems, where the interplay of gain and loss defies the constraints of traditional Hermitian frameworks \cite{Ashida2020,Bender1998,Bender2007,Heiss2004,El-Ganainy2007,Moiseyev2011,Gong2019}. A hallmark of this field is the non-Hermitian skin effect (NHSE), where a macroscopic number of eigenstates localize at the boundaries under open boundary conditions, challenging the conventional bulk-boundary correspondence \cite{Zhang2022Review,BOBBC_Kunst2018,Skin_Lee2016,Skin_Xiong2018,Skin_Yao2018}. This phenomenon, driven by non-reciprocal dynamics, necessitates novel theoretical constructs like the generalized Brillouin zone \cite{Yang2020GBZ,Yao2018,Yokomizo2019GBZ,Zhang2020GBZ}. Recent experimental breakthroughs have brought the NHSE into sharp focus, with observations in acoustic metamaterials \cite{WangMetaMaterials,Zhang2021Nature,Wen2022MetaM}, photonic lattices \cite{Xiao2020Photonic,Weidemann2020,Wang2025photonic}, and mechanical systems \cite{Scheibner2020Elastic,Xue2022Accoustics,Li2024Mechanical}, underscoring its relevance across diverse physical platforms—yet these realizations have largely been in non-interacting regimes.

Building on this foundation, the role of electron interactions in shaping the NHSE has emerged as a burgeoning frontier, particularly in strongly correlated systems where interactions drive phenomena like Mott insulation and magnetism \cite{Hamazaki2019,Liu2020,Lee2020MBNHSE,Hayata2021,Yoshida2021,Dora2022,Kawabata2022MB,Zhang2022,Faugno2022,Faisal2022,Wang2023,Longhi2023,Hayata2023,DiogoSoares2024,YoshidaNHMott,Sticlet2024,Kai2025,Brighi2024,Kim2024,Wan2025,Ling2025,Longhi2025,Zhang2025,Ibarra_Garc_a_Padilla2025,Qin2025}. The Hubbard model, extended to non-Hermitian settings with asymmetric hopping, serves as an ideal platform to probe how interactions suppress or modify boundary localization. Recent theoretical efforts illustrate this interplay: for example, Faber polynomial simulations of non-unitary dynamics in interacting Hatano-Nelson models reveal that interactions preserve antiferromagnetic order against skin accumulation at strong couplings \cite{DiogoSoares2024}. Similarly, autoregressive neural quantum states applied to the Hatano-Nelson-Hubbard model highlight optimization challenges under strong non-Hermiticity, mitigated by parameter ramping but exacerbated at large asymmetries \cite{Ibarra_Garc_a_Padilla2025}. Furthermore, the Pauli exclusion principle in fermionic systems suppresses NHSE by preventing over-accumulation at boundaries, enforcing electron spreading and favoring bulk phases—especially when combined with interaction-induced scattering that stabilizes delocalized states \cite{Brighi2024,Orito2023}.

However, these insights are often constrained by computational limitations, such as exact diagonalization's exponential scaling, which restricts analyses to small systems \cite{Hamazaki2019,Liu2020, Lee2020MBNHSE,Faisal2022,Brighi2024,Kim2024,DiogoSoares2024}. To overcome this and capture the intricate competition between correlations and spatial inhomogeneities in non-Hermitian lattices, scalable tools are essential. Among them, dynamical mean-field theory (DMFT) stands out as a computationally feasible treatment of local interactions within a self-consistent framework, yielding insights into complex phase diagrams in strongly correlated regimes~\cite{Metzner1989DMFT,Georges1996}. Its recent application to non-Hermitian physics has unveiled topological features like bulk Fermi arcs in heavy-fermion systems \cite{Nagai2020DMFT}, while the real-space extension (R-DMFT)—well-established for lower-dimensional systems with its ability to handle spatial inhomogeneity \cite{Dobrosavljevic1998,Freericks2004,Zgid2019,Pothoff1999,Helmes2008,Okamoto2004,Miranda2012}, is particularly well-suited for studying the NHSE in correlated systems, as demonstrated in recent analyses \cite{Yoshida2021,PetersYoshida2024RDMFT}. These analyses typically employ pseudo-spectrum techniques, deriving effective non-Hermitian Hamiltonians from self-energies computed in correlated Hermitian models to explore topology and correlation induced skin effects.

In this work, we generalize the R-DMFT to directly study non-Hermitian many-body systems, specifically the non-Hermitian Hubbard model at half-filling with asymmetric hopping terms that induce the NHSE. This approach addresses the computational challenge of incorporating both strong electron correlations and spatially dependent non-Hermitian effects on a lattice and also allows us to study their intricate interplay. We systematically investigate the competition between correlation-driven phenomena and boundary localization characteristic of the NHSE. While electron interactions typically drive bulk localization, the NHSE pushes states toward boundaries. We attempt to address the question of which mechanism dominates, and under what conditions? Our key finding reveals that at moderate non-Hermitian asymmetry, repulsive interactions suppress NHSE features, favoring correlation-driven bulk phases like Mott insulators. However, sufficiently strong non-Hermitian asymmetry can override correlation-induced suppression, restoring pronounced NHSE signatures even under strong electron correlations.
 
The structure of this paper is as follows: In Section ~\hyperref[Subsection:Model]{II A}, we define the non-Hermitian Hubbard model with asymmetric hopping. Section~\hyperref[Subsection:Method]{II B}  details the R-DMFT methodology employed in our study. In Section ~\hyperref[Subsection:Tool]{II C}, we briefly discuss the definition of end-to-end Green's function. In Section ~\hyperref[Section:Results]{III}, we present our results on the spectral function and Green's function analyses, discussing the implications for the skin effect. Finally, Section ~\hyperref[Section:Conclusion]{IV} concludes with a summary of our findings and potential directions for future research.

\section{Model and method}
\subsection{Non-Hermitian Hubbard Model}\label{Subsection:Model}
We consider the non-Hermitian Hubbard model on a one-dimensional lattice with $N$ sites under open boundary conditions (OBC), defined by the Hamiltonian~\cite{Hayata2021,Hatano1996,Hatano1998,HatanoNelson1998,HatanoNelson1997}
\begin{equation}
H = -\sum_{j,\sigma} \left( t_r c_{j+1,\sigma}^\dagger c_{j,\sigma} + t_l c_{j,\sigma}^\dagger c_{j+1,\sigma} \right) + U \sum_j n_{j\uparrow} n_{j\downarrow},
\label{eq:hamiltonian}
\end{equation}
where $c_{j,\sigma}^\dagger$ and $c_{j,\sigma}$ are the fermionic creation and annihilation operators at site $j$ with spin $\sigma = \uparrow, \downarrow$, and $n_{j\sigma} = c_{j,\sigma}^\dagger c_{j,\sigma}$ is the number operator. The hopping amplitudes are asymmetric, with $t_r = t + \gamma$ and $t_l = t - \gamma$, where $t$ is the base hopping strength and $\gamma \neq 0$ introduces non-Hermiticity, modeling effects like biased particle flow or dissipation. 
We set $t=1$ and it serves as the energy scale.
The parameter $U$ represents the on-site Hubbard interaction strength, capturing strong electron correlations. We set the chemical potential to zero to ensure half-filling. The OBC ensure that the NHSE, characterized by eigenstate localization at the lattice boundaries, can be observed, as the absence of periodic connections enhances boundary effects. The non-Hermiticity breaks symmetries including the translation symmetry, the parity or reflection symmetry, but it possesses the particle-hole symmetry, time reversal symmetry, and two continuous symmetries: U(1) symmetry for charge conservation and SU(2) symmetry for spin conservation remain intact. 
For the half-filled case without imbalance hopping, the Hubbard model has a hidden SU(2) symmetry from the $\eta$-pairing, the imbalance hopping breaks this symmetry \cite{yang1990so,Yang1989,pernici1990spin}. For the OBC case, all of its eigenvalues are real for the non-interacting case even without the PT symmetry. One can show this by the similarity transform of the Hamiltonian, this remains valid even with the interaction $U$ as long as $|t| >|\gamma|$.

\subsection{Real-space DMFT}\label{Subsection:Method}
To investigate the interplay between strong correlations and the non-Hermitian
skin effect in the Hubbard model, we employ real-space dynamical mean-field theory (R-DMFT), also known as statistical DMFT and inhomogeneous DMFT \cite{Freericks2004,Zgid2019,Pothoff1999,Helmes2008,Okamoto2004,Miranda2012}. R-DMFT extends conventional DMFT to account for spatial inhomogeneities, making it ideal for systems with open boundary conditions where the non-Hermitian skin effect leads to site-dependent properties. This is achieved by incorporating local self-energies dependent on the lattice site, $\mathbf{\Sigma}_{ij,\sigma}= \Sigma_{i,\sigma} \delta_{ij}$, where $\delta_{ij}$ is the Kronecker delta function. These self-energies are computed by mapping the lattice problem onto a set of single-site impurity problems, each coupled to a self-consistent dynamical bath. The impurity model is described by the local effective action given by:
\begin{gather}
    \mathcal{S}_i^{eff} = -\int_0^{\beta} d\tau\int_0^{\beta} d\tau' \sum_\sigma c^\dagger_{i,\sigma}(\tau) \mathcal{G}_{i,\sigma}(\tau-\tau')^{-1} c_{i,\sigma}(\tau')  \nonumber \\
    + U\int_0^{\beta} d\tau  n_{i,\uparrow}(\tau)n_{i,\downarrow}(\tau),
    \label{eqn:Eff_action}
\end{gather}
where $i$ is the site index, $\tau$ is imaginary time, and $\mathcal{G}_{i,\sigma}(\tau-\tau')$ is the Weiss field, representing the dynamical mean field that accounts for the influence of all other lattice sites. In the R-DMFT algorithm, this dynamical mean field is determined in a self-consistent manner. Firstly, the site-dependent self-energy is obtained by solving the effective action in \eqref{eqn:Eff_action} using an impurity solver. The interacting lattice Green's function is computed via the lattice Dyson equation:
\begin{equation}
    \mathbf{G}_\sigma(\omega)^{-1} = \mathbf{G}^{(0)}_{\sigma}(\omega)^{-1} - \mathbf{\Sigma}_{\sigma}(\omega) \label{eqn:LattDyson},
\end{equation}
where, $\omega$ denotes the real frequency. The non-interacting lattice Green's function $\mathbf{G}^{(0)}_{\sigma}(\omega)$ in real space representation is given by 
\begin{equation}
    \mathbf{G}^{(0)}_{\sigma}(\omega)^{-1} = \omega\mathbf{I} - \mathbf{J},
\end{equation}
where $\mathbf{I}$ denotes the identity matrix and $\mathbf{J}$ denotes the matrix of hopping amplitudes. Secondly, the diagonal elements of the interacting lattice Green's function are identified as the local Green's function, i.e., $G_{i,\sigma}(\omega) = \mathbf{G}_{ii,\sigma}(\omega)$. To complete the self-consistent loop, the local Dyson equation is used to compute the dynamical mean field:
\begin{equation}
    \mathcal{G}_{i,\sigma}(\omega)^{-1} =  G_{i,\sigma}(\omega)^{-1} + \Sigma_{i,\sigma}(\omega).
\end{equation}
In order to compute the self-energy of the action in \eqref{eqn:Eff_action}, we employ an iterative perturbation theory (IPT) solver \cite{Kajueter1996}.

\subsection{End-to-End Green's function}\label{Subsection:Tool}

Our key quantity of interest is the lattice Green's function, which encodes information about directional amplification in non-Hermitian systems. For the non-interacting case, the Green's function is defined as: 
\begin{equation}
    \mathbf{G}^{(0)}_{\sigma}(\omega) = \frac{1}{\omega \mathbf{I} - \mathbf{J}}.
\end{equation}
Under open boundary conditions (OBC), the matrix $\omega \mathbf{I} - \mathbf{J}$ for a one-dimensional chain of length $N$ takes the form:
\begin{equation}
    \omega \mathbf{I} - \mathbf{J} = \begin{pmatrix}
    \omega & t_l & 0 & \ldots & \ldots &0 \\
    t_r & \omega & t_l & 0 & \ldots & 0 \\
    0 & t_r & \omega & t_l & \ldots & 0 \\
    \vdots & \ddots & \ddots & \ddots & \ddots & \vdots \\
    0 & \ldots & 0 & t_r & \omega & t_l \\ 
    0 & \ldots & \ldots & \ldots & t_r & \omega
    \end{pmatrix}_{N \times N}
\end{equation}
The elements $\mathbf{G}^{(0)}_{ij,\sigma}(\omega)$ represent the amplitude of propagation from site $j$ to site $i$ at frequency $\omega$. The end-to-end Green functions, $\mathbf{G}^{(0)}_{1N}(\omega)$ and $\mathbf{G}^{(0)}_{N1}(\omega)$, specifically connect the first and last sites of the chain. These can be expressed through the spectral representation:
\begin{equation}
    \mathbf{G}^{(0)}_{ij,\sigma}(\omega) = \sum_{n} \frac{\braket{i}{\psi^R_n}\braket{\psi^L_n}{j}}{\omega - E_n}, \label{eqn:spect_rep}
\end{equation}
where $\ket{\psi^R_n}$ and $\ket{\psi^L_n}$ are the right and left eigenvectors of $\mathbf{J}$, respectively, with eigenvalues $E_n$, satisfying the bi-orthogonal normalization condition $\braket{\psi^L_m}{\psi^R_n} = \delta_{mn}$. 

The end-to-end Green's functions reveal directional amplification, a hallmark of non-Hermiticity~\cite{Wanjura2020,Yifei2020ChiralTunneling,Zirnstein2021BulkG,Li2022,Hu2023Greens,Huang2025DisGreens}. Specifically, when $|\mathbf{G}^{(0)}_{ij,\sigma}(\omega)| \gg 1$ while $|\mathbf{G}^{(0)}_{ji,\sigma}(\omega)| \ll 1$ for a certain pair
$(i,j)$, it indicates that a signal at frequency $\omega$ is amplified from site $j$ to site $i$, with suppressed back-propagation from $i$ to $j$ . In the Hatano Nelson model with $t_l > t_r$, under OBC, the non-Hermitian skin effect (NHSE) localizes all eigenstates at the left end of the chain. Consequently, the numerator $\langle 1|\psi^R_n\rangle\langle\psi^L_n|N\rangle$ becomes large for states localized at opposite boundaries. As a result, $\mathbf{G}_{1N,\sigma}(\omega)$ exhibits pronounced peaks when $\omega$ approaches the boundary‑localized eigenvalues, signaling the onset of the skin effect. In this work, we extend this analysis to the non-Hermitian Hubbard model, using the end-to-end Green's function to explore how onsite Hubbard interactions influence NHSE signatures and directional amplification.

\section{Results}\label{Section:Results}
\begin{figure}[tbp]
    \centering
    \includegraphics[width=\linewidth]{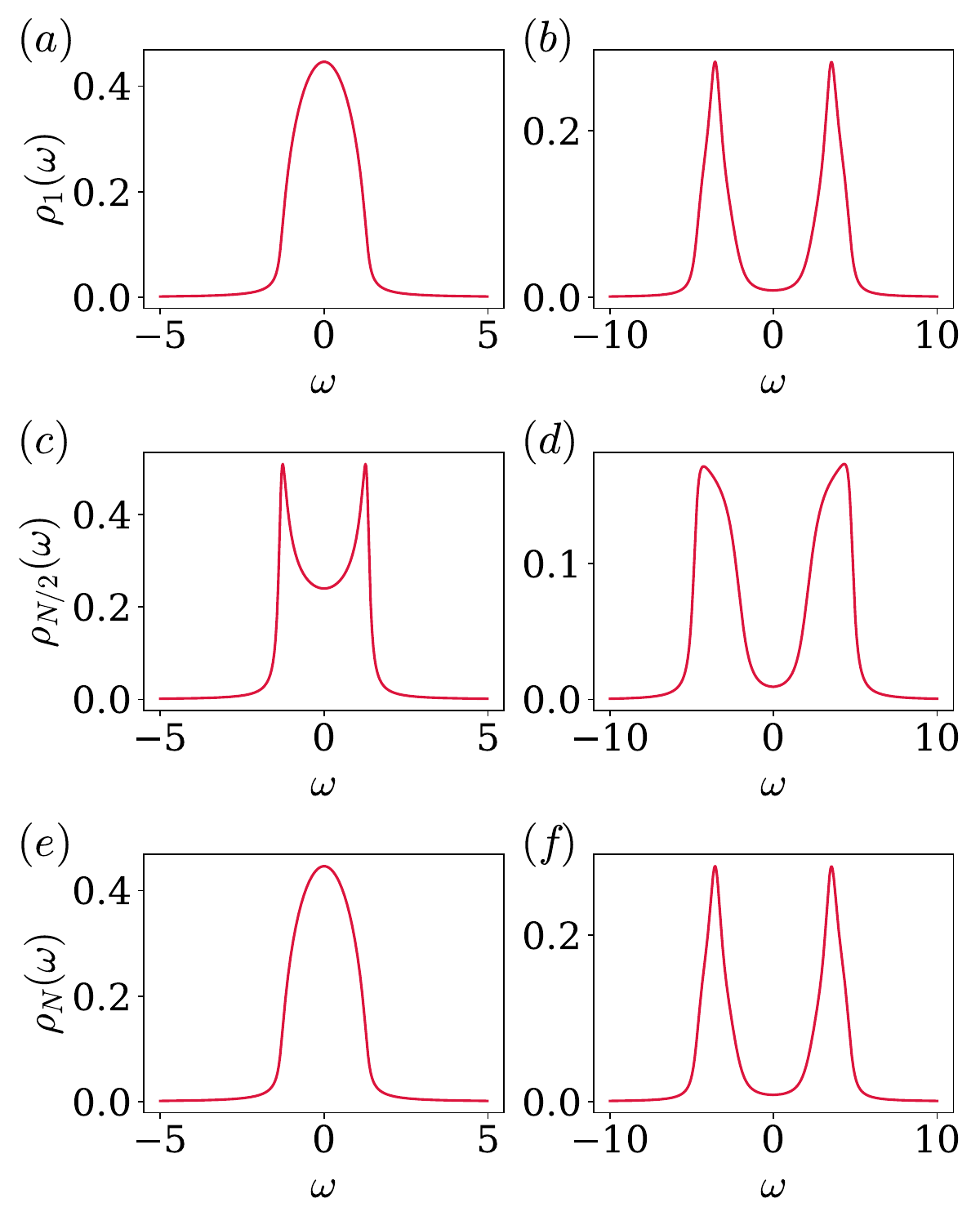}
    \caption{ Spectral function $\rho_i(\omega)$ [Eqn. \eqref{eqn:spect_func}] at various lattice sites for the 1D non-Hermitian Hubbard model. Panels (a), (c), and (e) correspond to the non-interacting case ($U = 0$) with non-Hermitian parameter $\gamma = 0.75$, while panels (b), (d), and (f) show the interacting case with $U = 5.0$ and $\gamma = 0.75$. The top row (a, b) presents results for the first site ($i = 1$), the middle row (c, d) for the central site ($i = N/2$), and the bottom row (e, f) for the last site ($i = N$).
}
    \label{fig:DOS}
\end{figure}

In this section, we proceed by investigating the 1D non-Hermitian Hubbard model at thermal equilibrium at temperature $T = 0.05$ using the R-DMFT algorithm. In our numerical simulations, we focus
on the paramagnetic solution and drop the spin index $\sigma$
for readability. We use a frequency grid size of $\Delta\omega=0.01$ for our R-DMFT simulations and IPT impurity solver. Furthermore, we use a small value of the spectral broadening factor $\eta = 0.1$ to keep our finite-size numerical simulations stable.  
The first quantity that we shall analyze is the spectral function $\rho_i(\omega)$ defined as follows:
\begin{equation}
    \rho_i(\omega) = -\frac{1}{\pi}\text{Im}[\mathbf{G}^R_{ii}(\omega+i\eta)]. \label{eqn:spect_func}
\end{equation}
Here, $\mathbf{G}^R_{ii}(\omega+i\eta)$ denotes the retarded component of the lattice Green's function $\mathbf{G}_{ij}(\omega)$, which is computed from our R-DMFT simulations. 

In Fig.~\ref{fig:DOS}, we present the spectral function $\rho_i(\omega)$ for various lattice sites of the 1D non-Hermitian Hubbard model. Figures \hyperref[fig:DOS]{1(a)}, \hyperref[fig:DOS]{1(c)}, and \hyperref[fig:DOS]{1(e)} illustrate the non-interacting limit ($U=0$) at a fixed non-Hermitian parameter $\gamma = 0.75$, while Figs. \hyperref[fig:DOS]{1(b)}, \hyperref[fig:DOS]{1(d)}, and \hyperref[fig:DOS]{1(f)} depict the interacting case with the same $\gamma = 0.75$ and an on-site interaction strength of $U = 5.0$. Specifically, the top row [Figs. \hyperref[fig:DOS]{1(a)} and \hyperref[fig:DOS]{1(b)}] shows the spectral function at the first lattice site ($i = 1$), the middle row [Figs. \hyperref[fig:DOS]{1(c)} and \hyperref[fig:DOS]{1(d)}] corresponds to the middle lattice site ($i = N/2$), and the bottom row [Figs. \hyperref[fig:DOS]{1(e)} and \hyperref[fig:DOS]{1(f)}] represents the spectral function at the last lattice site ($i = N$). In the interacting case, the spectral function reveals a gap opening around $\omega = 0$ and the presence of Hubbard bands. Another notable observation is that there is no difference or asymmetry in $\rho_i(\omega)$ between the first and last lattice sites of the system. Although the non-interacting limit exhibits a non-Hermitian skin effect with eigenstates localized at the left end of the chain, the spectral function alone does not clearly reflect this localization difference \cite{Yifei2020ChiralTunneling,Okuma2021NHSkin}. 
\begin{widetext}

    \begin{figure}[tbp]
        \centering
        \begin{minipage}{0.33\columnwidth}
            \centering
            \includegraphics[width=\columnwidth]{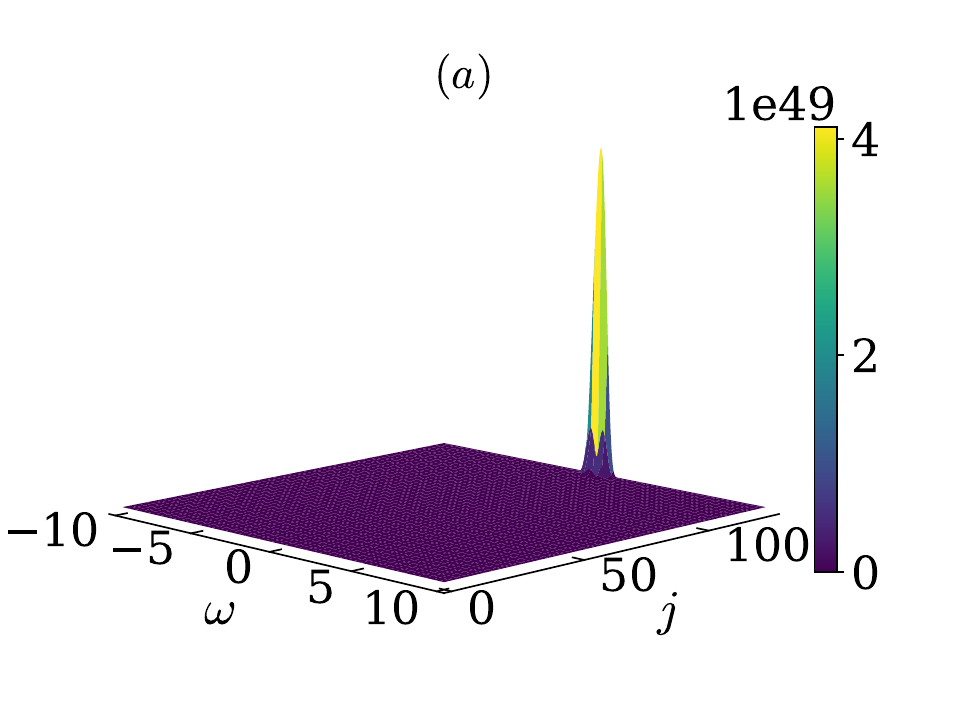} 
            \label{fig:3dPlotU0.0}
        \end{minipage}\hfill
        \begin{minipage}{0.33\columnwidth}
            \centering
            \includegraphics[width=\columnwidth]{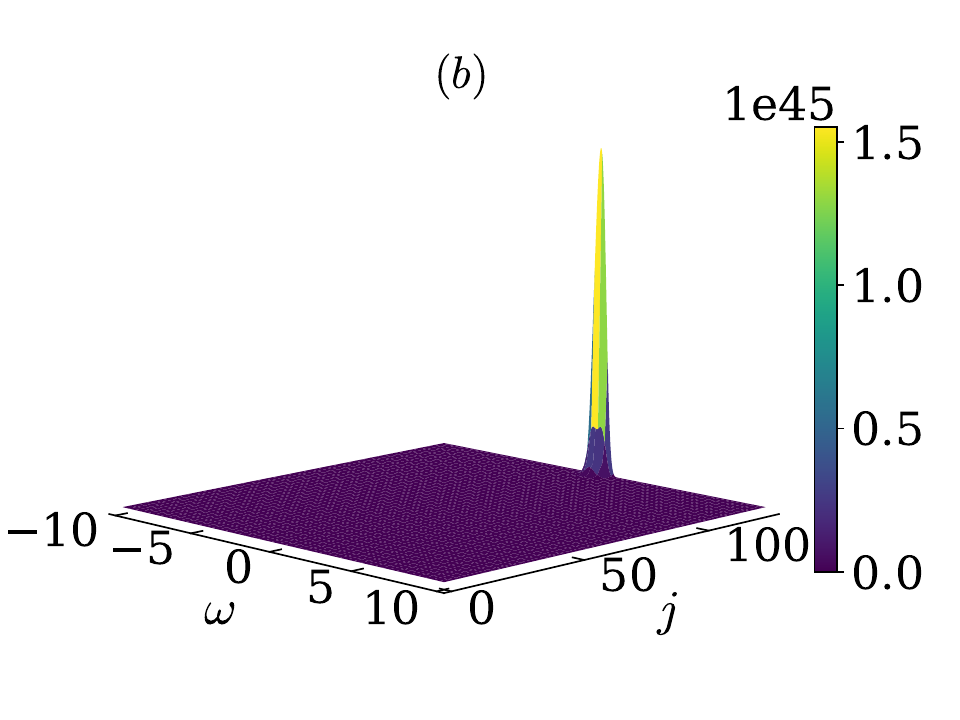} 
            \label{fig:3dPlotU0.5}
        \end{minipage}
        \begin{minipage}{0.33\columnwidth}
            \centering
        \includegraphics[width=\columnwidth]{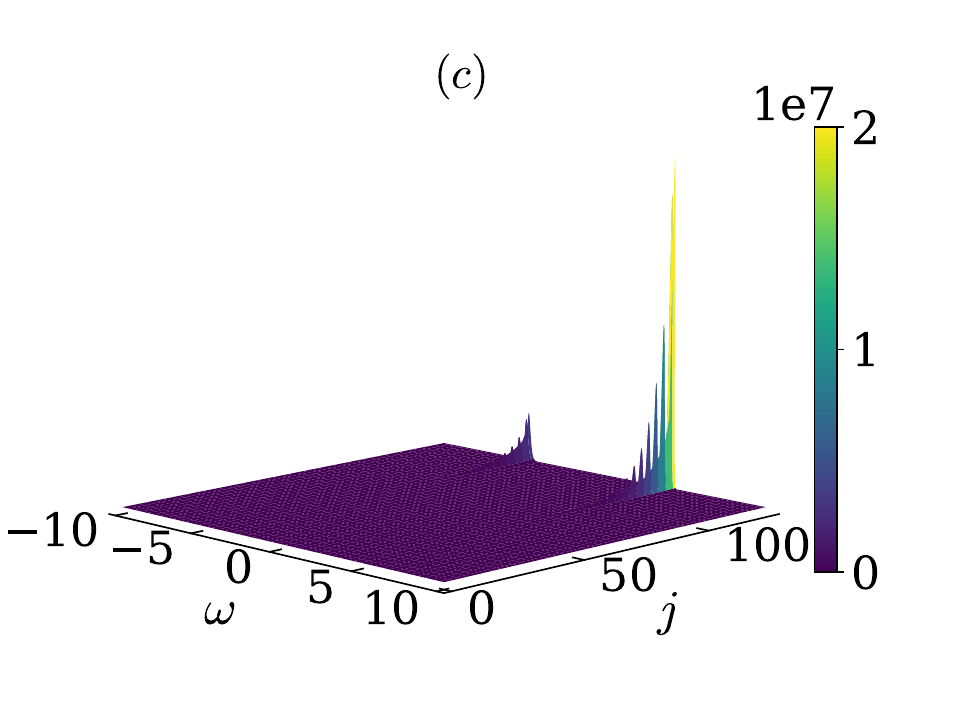}
        \label{fig:3dPlotU5.0}
        \end{minipage}
        \caption{$|\mathbf{G}_{1j}(\omega)|$ as a function of frequency $\omega$ and the site index $j$ for a non-hermitian strength of $\gamma = 0.75$ and various values of the Hubbard interaction (a) $U=0.0$ (non-interacting limit) (b) $U=0.5$ and (c) $U=5.0$. } 
        \label{fig:3dG1N}
    \end{figure}
    
\end{widetext}
To investigate the directional amplification induced by non-Hermiticity and its interplay with on-site Hubbard interactions, we examine the end-to-end Green's function $\mathbf{G}_{1j}(\omega)$. In Fig.~\ref{fig:3dG1N}, we present three-dimensional surface plots of $\mathbf{G}_{1j}(\omega)$ as a function of lattice index $j$ and frequency $\omega$, for a fixed non-Hermiticity parameter $\gamma = 0.75$ and three different interaction strengths: (a) $U=0$, (b) $U=0.5$, and (c) $U=5$. A key finding is that $\mathbf{G}_{1j}(\omega)$ peaks at the last site, $j = N$, most noticeably when $U = 0$. In this non-interacting case [Fig.~\hyperref[fig:3dPlotU0.0]{2(a)}], a sharp peak at $\omega = 0$ highlights the non-Hermitian skin effect, where particles accumulate at the chain’s edges due to asymmetric hopping \cite{Li2022ExactFormulaGreens}. For weak interactions ($U=0.5$), depicted in Fig.~\hyperref[fig:3dPlotU0.5]{2(b)}, the peak remains robust, although its magnitude is reduced by a few orders of magnitude. In the strongly interacting regime, illustrated in Fig.~\hyperref[fig:3dPlotU5.0]{2(c)}, the peak is further diminished by several orders of magnitude. Additionally, the interaction induces a shift in the peak positions, reflecting the redistribution of spectral weight towards higher-energy states. Moreover, the decay of $\mathbf{G}_{1j}(\omega)$ with $j$ is slightly slower than in the previous cases. These observations suggest that interactions reduce the directional amplification and redistribute spectral weight, possibly competing with the skin effect. To extend this analysis, we conduct a systematic study of end-to-end Green's function across a broader range of parameters. Specifically, we track the peak of the end-to-end Green's function for various non-hermitian parameter $\gamma$ and Hubbard interaction strengths $U$. We define $\omega_{ peak}$ as the frequency corresponding to this peak, which is typically near $\omega = 0$ in the absence of interactions.

\begin{figure}[tbp]
    \centering
    \includegraphics[width=\columnwidth]{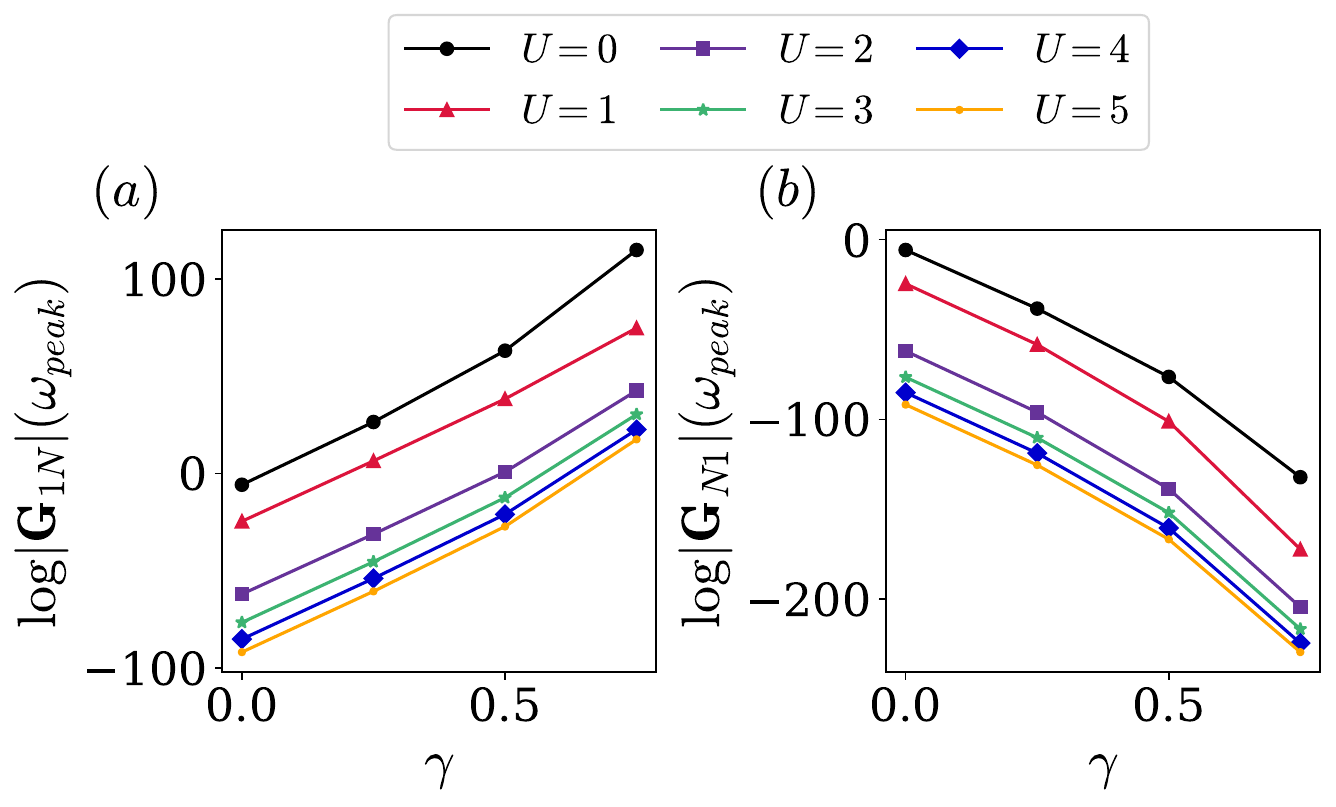}
    \caption{ Panels (a) and (b) display $\log | \mathbf{G}_{1N}(\omega_{peak}) |$ and $\log | \mathbf{G}_{N1}(\omega_{peak}) |$, respectively, as a function of the non-Hermitian asymmetry parameter $\gamma$ for several Hubbard interaction strengths $U$, with a fixed system size $N = 128$. }
    \label{fig:logG1NvsGamma}
\end{figure}

In Fig. \ref{fig:logG1NvsGamma}, we examine the end-to-end Green's functions $\log|\mathbf{G}_{1N}(\omega_{peak})|$ and $\log|\mathbf{G}_{N1}(\omega_{peak})|$ as a function of the non-Hermiticity parameter $\gamma$ for various values of the Hubbard interaction strengths with a system size fixed at $N = 128$. From Fig.~\hyperref[fig:logG1NvsGamma]{3(a)}, we observe that $\log |G_{1N}(\omega_{peak})|$ in the non-interacting case ($U = 0$) exhibits a pronounced increase with $\gamma$, a hallmark of the non-Hermitian skin effect, enhancing the propagation amplitude from site $N$ to site $1$. However, as interaction strength increases to finite values ($U > 0$), $\log |G_{1N}(\omega_{peak})|$ decreases relative to the non-interacting scenario for the same $\gamma$, indicating that electron-electron interactions suppress this directional amplification effect. This suppression becomes more evident with larger $U$, suggesting that many-body correlations compete with non-Hermiticity, thereby diminishing the directional amplification associated with the skin effect and favoring bulk-like behavior over boundary localization. In contrast, $\log |\mathbf{G}_{N1}(\omega_{peak})|$ [Fig.~\hyperref[fig:logG1NvsGamma]{3(b)}], which quantifies propagation from site $1$ to site $N$, decreases as the non-Hermitian asymmetry $\gamma$ grows, reflecting enhanced leftward hopping relative to rightward hopping ($t_l>t_r$). This pattern persists across all interaction strengths examined. The increase in the Hubbard interaction tends to suppress this further as evident by the decrease in $\log |\mathbf{G}_{N1}(\omega_{peak})|$ relative to the non-interacting scenario for the same $\gamma$. However, an intriguing observation emerges: at higher $\gamma$, the non-Hermitian asymmetry competes with the suppression induced by strong electron-electron interactions, as illustrated by the curve for $U=5$ (yellow line) in Fig.~\ref{fig:logG1NvsGamma}(a), which pushes the values of $\log |G_{1N}(\omega_{peak})|$ into the positive regime.

\begin{figure}[bp]
    \centering
    \includegraphics[width=\columnwidth]{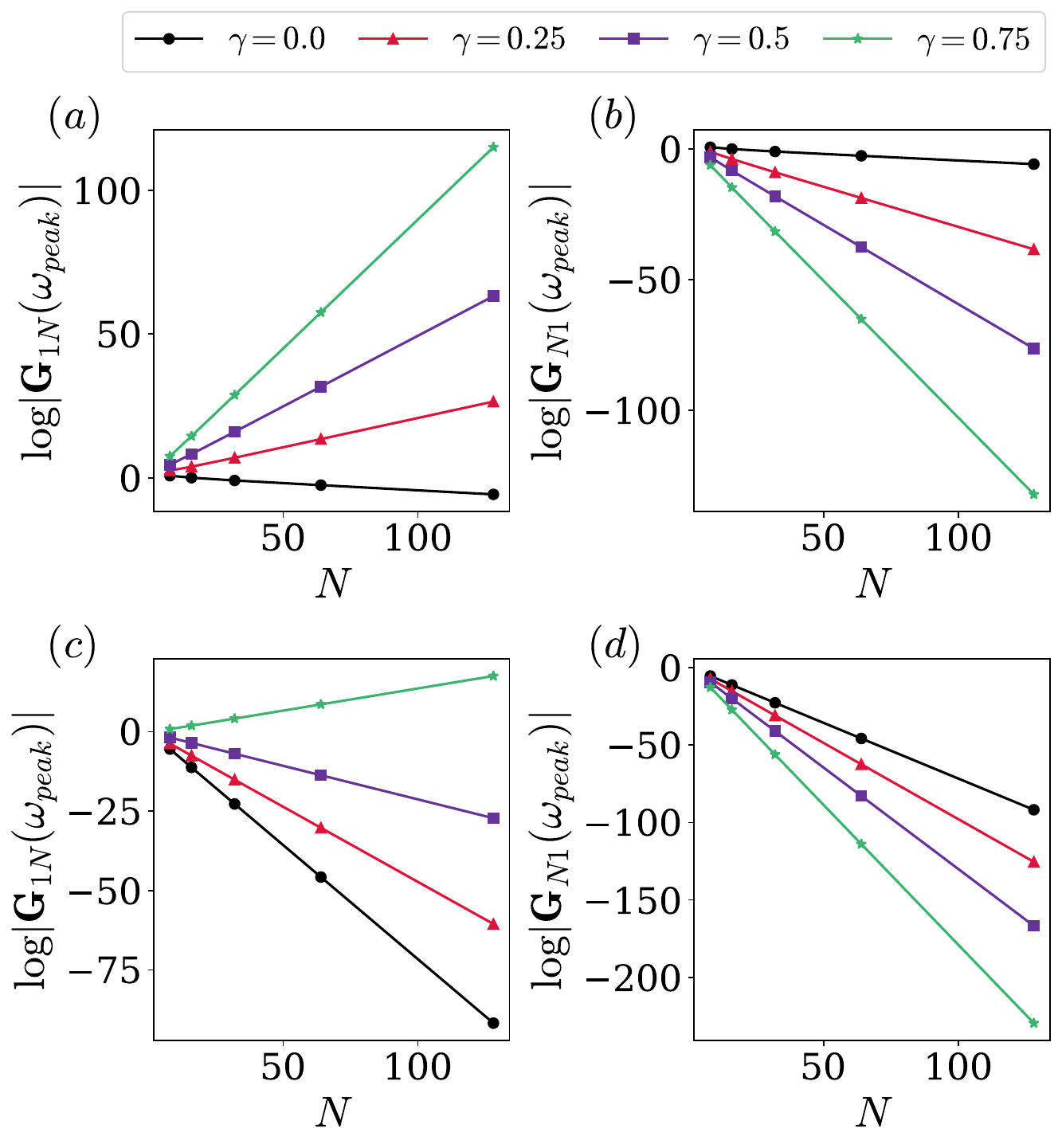}
    \caption{Panels (a) and (b) display the scaling of the end-to-end Green's function with system size, specifically $\log | \mathbf{G}_{1N}(\omega_{peak}) |$ and $\log | \mathbf{G}_{N1}(\omega_{peak}) |$, respectively, as a function of system size $N$ for various non-Hermitian strength parameter $\gamma$ in the non-interacting limit ($U=0$). Panels (c) and (d) show the scaling behavior with interactions turned on ($U = 5$).}
    \label{fig:scalingG}
\end{figure}

We further study the scaling of the end-to-end Green's functions as the system size is varied. Figure~\ref{fig:scalingG} illustrates the scaling behavior of the end-to-end Green's function with system size $N$. Panels (a) and (b) depict $\log | \mathbf{G}_{1N}(\omega_{peak})|$ and $\log | \mathbf{G}_{N1}(\omega_{peak}) |$, respectively, as a function of $N$ for various non-Hermitian strength parameters $\gamma$ in the non-interacting limit ($U = 0$). In Fig.~\hyperref[fig:scalingG]{4(a)}, $\log | \mathbf{G}_{1N}(\omega_{ peak}) |$ exhibits a clear linear increase with $N$ for finite $\gamma > 0$, indicative of exponential growth in $| \mathbf{G}_{1N}(\omega_{ peak}) |$. The slope of this growth steepens with larger $\gamma$, reflecting enhanced directional amplification from site $N$ to site $1$, a hallmark of the non-Hermitian skin effect (NHSE). Conversely, Fig.~\hyperref[fig:scalingG]{4(b)} shows a linear decrease in $\log | \mathbf{G}_{N1}(\omega_{ peak}) |$, corresponding to exponential decay in the reverse direction (from site $1$ to site $N$), with the decay rate also increasing with $\gamma$. This asymmetry between forward and backward propagation underscores the non-reciprocal nature of NHSE in the non-interacting regime, consistent with theoretical predictions for models like the Hatano-Nelson chain \cite{Li2022ExactFormulaGreens}.
\\
Figures ~\hyperref[fig:scalingG]{4(c)} and ~\hyperref[fig:scalingG]{4(d)} present the scaling of the end-to-end Green's function with interactions turned on ($U = 5$). The introduction of strong Hubbard interactions markedly alters the scaling. In Fig.~\hyperref[fig:scalingG]{4(c)}, the exponential growth of $\log |\mathbf{G}_{1N}(\omega_{peak})|$ observed for $U = 0$ is significantly suppressed, exhibiting a crossover to exponential decay for smaller to intermediate values of $\gamma$; the slopes are reduced across all $\gamma$, and for larger $\gamma$, the scaling approaches a plateau or weak linear dependence on $N$. However, as seen for $\gamma = 0.75$, a positive slope persists, suggesting that directional amplification remains robust and can dominate over strong electron-electron correlations in this regime. Overall, the attenuation for lower $\gamma$ contrasted with the resurgence at higher $\gamma$ highlights a nuanced competition: while many-body correlations tend to disrupt boundary localization through spectral weight redistribution to higher-energy Hubbard bands and scattering mechanisms that hinder directional amplification, sufficiently strong non-Hermitian asymmetry can prevail, allowing NHSE-driven effects to reassert themselves. Similarly, in Fig.~\hyperref[fig:scalingG]{4(d)}, the decay of $\log | \mathbf{G}_{N1}(\omega_{peak}) |$ is more pronounced compared to the non-interacting case, with steeper negative slopes.

These scaling trends provide compelling evidence of the interplay between non-Hermiticity and electron-electron interactions. In the non-interacting limit, the exponential scaling quantifies the robustness of NHSE, with growth/decay rates proportional to $\gamma$, reflecting the logarithmic amplification factor characteristic of skin modes. However, finite $U$ tempers this effect, as the reduced slopes suggest a crossover from boundary-dominated to correlation-driven dynamics, where Mott-like physics may favor uniformly localized states or suppress skin mode accumulation. This competition is particularly evident for small to intermediate $\gamma$, where the non-interacting amplification is strongest but most vulnerable to interaction effects; yet, for larger $\gamma$, the directional amplification can reassert itself and dominate over strong correlations, as seen in the persistent positive slopes.

\section{Conclusion}\label{Section:Conclusion}

We have generalized the real-space dynamical mean-field theory (R-DMFT) to investigate the intricate interplay between the non-Hermitian skin effect (NHSE) and electron-electron correlations in the 1D non-Hermitian Hubbard model with asymmetric hopping. By leveraging the non-Hermitian Green's function as a diagnostic tool—particularly the end-to-end components $\mathbf{G}_{1N,\sigma}(\omega)$ and $\mathbf{G}_{N1,\sigma}(\omega)$—we probed directional amplification, a hallmark of NHSE, since local spectral functions failed to reveal clear boundary asymmetries. In the non-interacting limit, $G_{1N}(\omega)$ exhibited exponential growth with system size $N$ and non-Hermitian asymmetry $\gamma$, confirming robust directional amplification and non-reciprocal propagation. However, finite interactions markedly suppressed this amplification, reducing peak magnitudes and shifting $\omega_{\rm peak}$ to finite frequencies aligned with Hubbard bands. Additionally, we observed changes in the scaling behaviors, signaling a crossover from boundary-dominated to correlation-driven dynamics, where many-body effects—such as Mott physics—redistribute spectral weight and potentially introduce scattering mechanisms that hinder directional amplification. A systematic parameter sweep over $\gamma$ and $U$ further revealed a nuanced competition: while correlations dominate for small to intermediate $\gamma$, leading to exponential decay in $\log |G_{1N}|$, higher $\gamma$ can reassert directional amplification, pushing values positive even at strong $U = 5$. These findings enhance our understanding of the intricate interplay between strong electron correlations and non-Hermitian physics, underscoring the potential for tunable non-reciprocal phenomena in correlated open quantum systems. The generic method employed here can be generalized to study the effects of correlations and random disorder, and extended to systems with large unit cells or non-periodic systems, such as quasicrystals \cite{Wernsdorfer2011,Mahmoudian2015,Tam2021}, which have recently garnered significant attention \cite{Longhi2021,Rangi2024,Shang2025,Zhang2024,Zhou2021,Suthar2022}.

\begin{acknowledgments}
This work used high-performance computational resources provided by the Louisiana Optical Network Initiative and HPC@LSU computing.
\end{acknowledgments}

\bibliography{references}
\newpage
\appendix
\onecolumngrid
\end{document}